\documentclass[showpacs,preprintnumbers,prd,nofootinbib,floats,amssymb,floatfix]{revtex4}

\usepackage{mathrsfs}
\usepackage{amsfonts}
\usepackage{amsmath}
\usepackage{amssymb}
\usepackage{dsfont}                              % fuente mathds
\usepackage{amsthm}
\usepackage{latexsym}                            % fuente para usar simbolos
\usepackage{bm}                                  % bold math
\usepackage[bbgreekl]{mathbbol}                  % fuetes griegas modificadas
\usepackage{hyperref}
\usepackage[retainorgcmds]{IEEEtrantools}        % eqnarray modificado

\newcommand{\nn}{\nonumber}
\newcommand{\yn}{\IEEEyessubnumber}

\begin{document}
%\begin{flushleft}
\title{ Hamiltonian analysis for  abelian and
non-abelian  massive theories in three dimensions}
%\end{flushleft}
\author{ Alberto Escalante } \email{aescalan@ifuap.buap.mx}
\author{ Jos\'e L. Osio}  \email{jalopez@ifuap.buap.mx}
%\author{ Moises Zarate}  \email{mzarate@ifuap.buap.mx}
\affiliation{
 Instituto de F{\'i}sica Luis Rivera Terrazas, Benem\'erita Universidad Aut\'onoma de Puebla, (IFUAP). Apartado
 postal J-48 72570 Puebla. Pue., M\'exico, 20 octubre 2009 }

%\maketitle

%\tableofcontents

\begin{abstract}
 A pure Dirac's method  for   abelian and non-abelian  massive theories    in three dimensions  is performed. Our  analysis  is developed on the extended phase space  reporting  the relevant structure of the theories,  namely, the extended    action, the extended Hamiltonian, the full structure of the constraints and the counting of degrees of freedom. In addition,  we compare our results with those   found in the literature.  

\end{abstract}

\date{\today}
\preprint{}
\maketitle

%%%%%%%%%%%%%%%%%%%%%%%%%%%%%%%%%%%%%%%%%%%%%%%%%%%%%%%%%%%%%%%%%%%%%%%%%%%%%%%%%%%%%%%%%%%%%%%%%%%%%%%%%%%%%%%%%%%%%%%%%%%%
\section{Introduction}
Topological field theories are characterized by being devoid of
local degrees of freedom. That is, the theories are susceptible only
to global degrees of freedom associated with non-trivial topologies
of the base manifold and  the
gauge bundle \cite{1, 2}. Examples of topological field theories, are  those called $BF$ theories.
These  theories were introduced as generalizations
of three-dimensional Chern-Simons actions or in other cases, can
also be considered as a zero coupling limit of Yang-Mills [YM] theories
\cite{3,4}. The importance for studying topological actions has been motived
in several contexts of theoretical physics given that  have  a
closed relation with physical theories,  for instance  it is well-know that from the  Chern-Simons topological action  arises a topological mass when it is added to actions just as electrodynamics or gravity  in  three dimensions \cite{ 5, 6};    at quantum level  an infrared cut-off  cures  the infrared problem without disturbing the ultraviolet or gauge properties of those theories.  On the other hand, in \cite{7} it   is reported an analysis of  specific limits in the gauge coupling of topological theories yielding a pure YM  dynamics in four and three  dimensions.  In this respect, in  the  four-dimensional case   nonperturbative topological configurations of the gauge fields is defined having a important role in realistic theories as for instance  quantum chromodynamics. Furthermore,  the  thee dimensional case is analyzed  at Lagrangian level, and the action  is  the coupling of $BF$-like  theories in order to generalize the quantum dynamics of YM and thus provide possible extensions to the quantum dynamics of three dimensional gravity. \\
With these antecedents in mind, the purpose of this paper is to report a pure Dirac's method of constrained
systems  for the three dimensional topologically massive  actions purposed in \cite{7}. The approach developed in this work,  is  quite different to the standard Dirac's analysis where usually  the dynamical variables are considered
only those that occurs in the action with temporal derivative. With the terminology ''a pure Dirac's
method''  we mean that  we will consider in the Hamiltonian framework that all the fields that define
our theory are dynamical ones;
this fact will allow us to find the complete structure of the constraints, the equations
of motion, gauge transformations, the extended action as well as the extended
Hamiltonian. Hence, in our approach we obtain results that   extend those reported in \cite{7}. It is important to remark that in the case of  three dimensional tetrad gravity, in despite of the existence of several  articles performing the hamiltonian analysis in the standard approach, there exist  problems in order to report  the correct gauge transformations of the theory \cite{8}; in some papers it is written that the gauge symmetry is Poincar\'e symmetry \cite{Witten}, in others that is Lorentz symmetry plus diffeomorphisms \cite{Carlip, 10a, 10b, 10c}, or that  there exist various ways to define the constraints leading to different gauge transformations \cite{11}. In fact,  in \cite{11} by using  variables defined on the spacetime,  the total  Hamiltonian does not have a direct geometrical meaning.  Hence, the constraints are replaced by equivalent  ones, however this fact  yield different  gauge symmetry. Furthermore, in 4D-gravity there exist  similar problems;  we are  able to realize  that developing the Hamiltonian approach on a smaller phase space context,   the structure obtained for  the constraints is not right \cite{12, 17, 17a}.  In fact,  we  observe in \cite{11b} that the Hamiltonian constraint for Palatini theory do not has the required structure  to   form a closed algebra with all  constraints;  this problem emerges because of  by working on a smaller phase space context  we  lose  control on the constraints,  and   to obtain  the correct structure sometimes they need to be fixed by hand.  We think that   a pure Dirac's formalism  is the best  tool  for solving these problems in order  to know the correct constraints  and therefore the correct  gauge symmetry. In this manner, we develop in this paper  all the Dirac's steps and we report a complete hamiltonian description of the models studied   in \cite{7}.\\
The paper is organized as follows: In Section II, we perform a pure Dirac's method for  the abelian massive  Govaerts's model, we develop a complete analysis and we report  the extended action, the full  constraints program, the gauge transformations and  the Dirac's brackets structure. In Section III, we extend the results found in previous sections  for a non-abelian theory; we report the extended action, the full constraints program and we perform the counting of physical degrees of freedom. In Section IV  we discuss on remarks and conclusions.
\section{Hamiltonian dynamics}\label{section:hd}
In this section, we will perform the hamiltonian analysis of the theory reported in \cite{7} given by 
\begin{equation}
S[A,\varphi] = \int_M \mathrm{d}^3 x \left\{ \frac{1}{2} \varepsilon^{\mu \nu \rho}
    \varphi_{\mu} F_{\nu \rho} + \frac{1}{2} e^{2} \varphi_{\mu} \varphi^{\mu} + \frac{1}{2} N
    \varepsilon^{\mu \nu \rho} A_{\mu} F_{\nu \rho} \right\} \,,
     \label{eq1}
\end{equation}
where  $e$ and $N$ are coupling  constants, $\varphi_{\mu}$ is a dynamical field,   the curvature tensor  $F_{\mu \nu}$ is defined as usual $F_{\mu \nu} = \partial_{\mu} A_{\nu} -\partial_{\nu} A_{\mu}$, the  $\mu, \nu=0,1,2$ are spacetime indices and $x^\mu$  are the coordinates that label the points for the three-dimensional Minkowski manifold $M$ with signature $\eta_{\mu \nu}= (-1,1,1)$.  As we have commented above, a complete  Hamiltonian analysis of the action (\ref{eq1}) has not been developed,  and becomes to be an  important step to understand the symmetries of the theory  in both  classical and quantum regimen. It is important to remark,  that is not common develop   a pure Dirac's method  for field theories (working with the
complete phase space). The principal reasons for studying the Hamiltonian formalism
under a smaller phase space context (standard Dirac's approach)  and not carried out on the complete phase space, is because
the separation of the constraints into first or second class is not easy to carry out. In this manner,
in the literature we find that the people prefer to work on a smaller phase space context because generally there are present only first class constraints and is common to avoid the difficult part of the
separation among the constraints in first and second class. The price that we pay by working  on a smaller phase space context is
that we can not neither know the complete form of the constraints, the complete form of gauge transformations
defined on the full space phase nor the complete algebra among the constraints for the theory
under study. For instance, for $BF$ theory there exist several studies developed on a smaller phase space context, however  in \cite{12} it has been reported  the complete structure of the constraints  on the full phase space by using a pure Dirac's approach.  Of course, by working with the full configuration space we are able to reproduce the results
obtained by working on a smaller configuration space. Hence,  it is mandatory to develop a complete Dirac's analysis  meaning that all steps of the Dirac formulation must be performed \cite{8}. It is well-know that if some of Dirac's steps  are  ignored or implemented incorrectly, we lose the control  of the relevant symmetries of the theory;  the constraints,   the extended action  and    gauge symmetry \cite{12}.   \\
The equations of motion obtained from the variation of (\ref{eq1}) are given by 
\begin{eqnarray}
\frac{1}{2} \epsilon^{\mu \nu \rho} F_{\nu \rho} + e^2 \varphi^{\mu}&=&0, \nonumber \\
\partial_\nu F^{\nu \mu} - \frac{Ne^2}{2} \star F^{\mu}&=&0,
\end{eqnarray}
where $\star F^{\mu} = \frac{1}{2} \epsilon^{\mu \nu \rho} F_{\nu \rho} $. Hence, by substituting the first equation of motion in the second, we obtain 
\begin{equation}
\left( \square + 2\left( Ne^2 \right)^2  \right) \star F^\mu=0, 
\end{equation}
 thus, the equations of motion represent a massive gauge field and there is not reference for the vector potential.  It is important to comment,  that in the limit $e \rightarrow0$ the action  (\ref{eq1}) becomes to be a topological one, this means, in the limit the theory is devoid  of physical degrees of freedom. In fact, the first term of  (\ref{eq1}) is  $BF$-type and the third is the conventional Chern-Simons action.  \\
 For our aims, we   perform  the 2+1 decomposition, thus,  the action (\ref{eq1}) takes the following form 
\begin{eqnarray}
S[A,\varphi]
    & = & \int \mathrm{d}^3 x \bigg\{ \frac{1}{2} \varepsilon^{\mu \nu \rho} \big( \varphi_{\mu} + N A_{\mu}
          \big) F_{\nu \rho} + \frac{1}{2} e^{2} \varphi_{\mu} \varphi^{\mu} \bigg\}                             \nn \\
    & = & \int \mathrm{d}^3 x \bigg\{ \frac{1}{2}\eta^{ i j}\varphi_0 F_{ij} + \frac{e^2}{2}\varphi_0 \varphi^0 + \dot{A}_{i}\eta^{ij}(\varphi_j + N A_j) +A_0\eta^{ij}(\partial_i \varphi_j + 2N\partial_iA_j)  \nonumber \\
    &+& \frac{e^2}{2} \varphi_i \varphi^i  \bigg\} \, ,
\end{eqnarray}
where $\varepsilon^{0ij}\equiv \eta^{ij}$. A pure Dirac's analysis calls  for the definition of the momenta $(\Pi^{\mu}, \pi^{\mu})$  canonically conjugate to ($A_{\mu}$, $\phi_{\mu}$) 
\begin{equation}
\label{eq:momentos}
 \left\{
       \begin{array}{r c l c l}
          \Pi^{0} & = & \frac{\partial \mathcal{L}}{\partial \dot{A}_{0}}       & = & 0,                         \\[0.5em]
          \Pi^{i} & = & \frac{\partial \mathcal{L}}{\partial \dot{A}_{i}}       & = &
                        \eta^{ i j} \big( \varphi_{j} + N A_{j} \big),                                    \\[0.5em]
          \pi^{0} & = & \frac{\partial \mathcal{L}}{\partial \dot{\varphi}_{0}} & = & 0,                          \\[0.5em]
          \pi^{i} & = & \frac{\partial \mathcal{L}}{\partial \dot{\varphi}_{i}} & = & 0.
       \end{array}
\right.
\end{equation}
The Hessian of the Lagrangian is given by 
\begin{equation}
\label{eq:hessiana}
\mathds{H}_{A B} \equiv \frac{ \partial^{2} L }{ \partial \dot{Q}_{A} \partial \dot{Q}_{B} }= 0 \, .
\end{equation}
where $Q_{A}$ and $P^{A}$ will  label  the sets of variables $\{ A_{\mu} , \phi_{\mu} \}$ and $\{ \Pi^{\mu} , \pi^{\mu} \}$ res\-pec\-tively.  The rank of the Hessian is zero,  therefore we expect six independent primary constraints given by 
\begin{equation}
\label{eq5}
\left\{
      \begin{array}{c c l}
         \Phi^{0}& = & \Pi^{0} \approx 0,                                                                    \\[0.1em]
         \Phi^{i} & = & \Pi^{i} - \eta^{ i j} ( \varphi_{j} + N A_{j} ) \approx 0,                    \\[0.1em]
         \phi^{0} & = & \pi^{0} \approx 0,                                                                    \\[0.1em]
         \phi^{i} & = & \pi^{i} \approx 0.
      \end{array} \right.
\end{equation}
In order to obtain independent primary constraints, they  must satisfy  the regularity conditions.  Such conditions are given by  the following  Jacobbian matrix 
\begin{equation}
\label{eq:jacobiana}
\frac{\partial \big( \Phi^{1}_{\alpha} , \phi^{1}_{\alpha} \big)}{\partial \big( Q_{A} , P^{A} \big)} =
\left(
  \begin{array}{cccccccccccc}
    0  &  0  &  0  &  0  &  0  &  0  &  1  &  0  &  0  &  0  &  0  &  0                                          \\
    0  &  0  & -N  &  0  &  0  & -1  &  0  &  1  &  0  &  0  &  0  &  0                                          \\
    0  &  N  &  0  &  0  &  1  &  0  &  0  &  0  &  1  &  0  &  0  &  0                                          \\
    0  &  0  &  0  &  0  &  0  &  0  &  0  &  0  &  0  &  1  &  0  &  0                                          \\
    0  &  0  &  0  &  0  &  0  &  0  &  0  &  0  &  0  &  0  &  1  &  0                                          \\
    0  &  0  &  0  &  0  &  0  &  0  &  0  &  0  &  0  &  0  &  0  &  1                                          \\
  \end{array}
\right) \,,
\end{equation}
which  has constant rank on the constraints  surface. In this manner, the regularity conditions are satisfied.\\
The canonical Hamiltonian is defined  by
\begin{eqnarray}
\label{eq7}
H_c & = & \! \int \mathrm{d}^2 x \left\{ \dot{Q}_{A} P^{A} - \mathcal{L} \right\} = \int \mathrm{d}^{2} x \left\{
        \dot{A}_{\alpha} \Pi^{\alpha} + \dot{\varphi}_\alpha \pi^\alpha - \mathcal{L} \right\}                                                              \nn \\
  & = & \! \int \mathrm{d}^{2} x \bigg\{ -\frac{1}{2} \varphi_{0} \eta^{ij}F_{ij} - \frac{e^2}{2} \varphi_0 \varphi^0 - A_0( \partial_i \Pi^i +  \frac{N}{2} \eta^{ij}F_{ij}) - \frac{e^2}{2}\varphi_i \varphi^i \bigg\},                                                                                 
  \end{eqnarray}
here, $F_{ij}=\partial_i A_j- \partial_j A_i$.  On the other hand,  the primary Hamiltonian of the theory is given  by 
\begin{equation}
\label{eq:H_total}
H_{P} (x) = H (x) + \int \mathrm{d}^{2} x( \lambda_{\alpha}  \Phi^{\alpha} + \beta_{\alpha}\phi^{\alpha} )
\end{equation}
where $\lambda_{\alpha}$ and $\beta_{\alpha}$ denotes a   set of Lagrange multipliers enforcing the primary  constraints $\{ \Phi^{\alpha} , \phi^{\alpha} \}$. \\
The fundamental Poisson brackets  of the theory are defined by 
\begin{eqnarray}
\big[ A_\alpha (x^{0},x), \Pi^{\beta} (x^{0},y) \big] &=& \delta^{\beta}_{\alpha} \, \delta^{2} (x-y), \nonumber \\
\big[ \varphi_\alpha (x^{0},x), \pi^{\beta} (x^{0},y) \big] &=& \delta^{\beta}_{\alpha} \, \delta^{2} (x-y). 
\label{eq9}
\end{eqnarray}
Now,  we need  identify if the  theory presents   secondary constraints. For this purpose, we compute the  6 $\times$ 6 matrix whose entries are the Poisson brackets among the primary constraints (\ref{eq5}) 
\begin{eqnarray}
\label{eq}
\big[  \Phi^{0} (x), \Phi^{0} (y)  \big] &=&  0, \qquad \qquad  \qquad \qquad     \big[  \Phi^{0} (x), \Phi^{i} (y) \big] =  0,          \nonumber \\                                        
\big[  \Phi^{0} (x), \phi^{0} (y)  \big] &=&  0, \qquad \qquad  \qquad \qquad   \big[ \Phi^{0} (x), \phi^{i} (y)  \big] = 0,  \nonumber \\                  
\big[  \phi^{i} (x), \phi^{j} (y)  \big] &=& 0, \qquad \qquad  \qquad \qquad  \big[ \Phi^{i} (x), \phi^{0} (y)  \big] = 0,  \nonumber \\                     
\big[  \Phi^{i} (x), \phi^{j} (y)  \big]  &=& - \eta^{ij} \delta^2(x-y),  \quad \quad     \big[ \Phi^{i} (x), \Phi^{j} (y)  \big] = -2N \eta^{ij} \delta^2(x-y),  \nonumber \\                                                        
\big[  \phi^{i} (x), \phi^{0} (y)  \big] & =& 0,                                                        
\end{eqnarray}
by calling $\Delta \equiv (\Phi^\alpha, \phi^\alpha)$, the matrix takes the form
\begin{eqnarray*}
\label{eq}
\big[ \Delta (x) , \Delta (y) \big] =
\left(
  \begin{array}{cccccc}
    0   &   0   &   0     &   0   &   0   &   0                                                                  \\
    0   &   0   & 0   &   0   &   0   & - 1                                                                  \\
    0   & 0   &   0     &   0   &   1   &   0                                                                  \\
    0   &   0   &   0     &   0   &   0   &   0                                                                  \\
    0   &   0   & 1     &   0   &   0   &   -2N                                                                  \\
    0   &   -1   &   0     &   0   &   2N   &   0
  \end{array}
\right)  \delta^2(x-y).
\end{eqnarray*}
This matrix has rank=4 and   2 null vectors, so we expect 2 secondary constraints. In fact, from the temporal evolution of the constraints (\ref{eq5}) and the contraction with the 2 null vectors, it follows that  the following  2 secondary constraints arise
\begin{eqnarray}
    \dot{\phi}^{0}=[\phi^{0}(x), H_{P}(y) ]\approx 0  &\longrightarrow& \Lambda^0: e^2 \varphi^0 +\frac{1}{2}\eta^{ij} F_{ij}\approx 0,         \\
    \dot{\Phi}^{0}=[\Phi^{0}(x), H_{P}(y) ]\approx 0  &\longrightarrow& \Sigma^0: \frac{N}{2}\eta^{ i j} F_{ij} + \partial_{i} \Pi^{i} \approx 0,
\end{eqnarray}
and the rank allows  us to fix the following 4 Lagrange multipliers 
\begin{eqnarray}
 \dot{\Phi}^{i}= [\Phi^{i}(x), H_{P}(y) ]\approx 0 &\longrightarrow& \lambda_i = e^2 \eta_{li} \phi^l,\\
 \dot{\phi}^{i}= [\phi^{i}(x), H_{P}(y) ]\approx 0 &\longrightarrow& \beta_i = -2N e^2 \eta_{ij} \varphi^j - \partial_i \varphi^{0}.
 \end{eqnarray}
 For this theory there are not third constraints. In this manner,  we have found
\begin{equation*}
\centering
\begin{tabular}{l l}
Primary Constraints & $ \left\{
       \begin{array}{r c l}
          \Phi^{0} & = & \Pi^{0} \approx 0,                                                                   \\[0.1em]
          \Phi^{i} & = & \Pi^{i} - \eta^{ i j} ( \varphi_{j} + N A_{j} ) \approx 0,                   \\[0.1em]
          \phi^{0} & = & \pi^{0} \approx 0,                                                                   \\[0.1em]
          \phi^{i} & = & \pi^{i} \approx 0,
       \end{array} \right. $                                                                                     \\[2.5em]
Secondary Constraints & $ \left\{
       \begin{array}{r c l}
          \Lambda^0 & = & \frac{N}{2}\eta^{ i j} F_{ij}  +  \partial_{i} \Pi^{i}
                             \approx 0,                                                                     \\[0.1em]
          \Sigma^0 & = & e^{2} \varphi^{0} +\frac{1}{2} \eta^{i j} F_{ij} \approx 0,   \\[0.1em]
       \end{array} \right. $                                                                                     \\[1.1em]
Lagrange multipliers  $ \beta_{\alpha}, \lambda_{\alpha} $  & $ \left\{
       \begin{array}{r c l}
          \beta_{0} & = & \mathrm{unknown},                                                                 \\[0.1em]
          \lambda_i & = & e^{2} \eta_{li} \varphi^{l},                                                             \\[0.1em]
                   \lambda_{0}           & = &    \mathrm{unknown},                                                 \\[0.1em]
          \beta_{i}           & = & - \partial_{i} \varphi^{0} - 2 N e^{2}\eta_{ij} \varphi^{j}.                                 \\[0.1em]
       \end{array} \right. $
\end{tabular}
\end{equation*}
By following with the method, we need to separate from the primary and secondary constraints  which ones correspond to first and second class.  In order to archive  this aim, we need to calculate   the Poisson brackets  among   primary and secondary constraints which are given by the following 8$\times$8 matrix 
\small{\begin{equation*}
\label{eq:algebra_de_restricciones}
\bordermatrix{
                 & \Phi^{0} (y) & \Phi^{1} (y)  & \Phi^{2} (y) & \phi^{0} (y) & \phi^{1} (y)
                                              & \phi^{2} (y) & \Lambda^{0} (y) & \Sigma^{0} (y)  \cr
\Phi^{0} (x) & 0                & 0           & 0                & 0                & 0
                                              & 0                & 0                & 0           \cr
\Phi^{1} (x) & 0                & 0           & -2N\delta^2(x-y)             & 0                & 0
                                              & - \delta^2(x-y)               & \partial_{(x) 2} \delta^2(x-y)    & 0   \cr
\Phi^{2} (x) & 0               & 2N\delta^2(x-y)     & 0                & 0       & \delta^2(x-y) 
                               & 0                & -\partial_{(x) 1} \delta^2(x-y)     & 0      \cr
\phi^{0} (x) & 0                & 0                 & 0                & 0                & 0
                                              & 0               & e^{2}\delta^{2}(x-y)             & 0     \cr
\phi^{1} (x) & 0                & 0                 & -\delta^2(x-y)                & 0                & 0
                                              & 0                 & 0           & 0                          \cr
\phi^{2} (x) & 0                & \delta^2(x-y)                  & 0                & 0                & 0 
                                              & 0                & 0                & 0                 \cr
\Lambda^{0} (x) & 0                & -\partial_{(x) 2} \delta^2(x-y)                  & \partial_{(x) 1} \delta^2(x-y)    & -e^2\delta^2(x-y)                & 0
                                              & 0                & 0                & 0                          \cr
\Sigma^{0} (x) & 0                &0               &0                          & 0           & 0
                                              & 0                & 0                & 0                          }
\end{equation*}}

This matrix has rank=6 and nullity=2, consequently there are  2 first class
constraints $(\Phi^{0}, \Sigma^{0}) $ and  6 second class constraints  $( \phi^{0},\Phi^{i}, \phi^{i}, \Lambda^{0}) $. Because of there are six second class constraints, this implies that  must  be identified   six  Lagrange multipliers,  and we will work out in later lines on this step. In review, we have found \\
\begin{tabular}{|l|l|l|}
  \hline
    \: $1^{st}$ Class Constraints= 2
  & \: $2^{nd}$ Class Constraints=6
  & \: Lagrange Multipliers                                                                                      \\
  \hline \hline
\: $\gamma^{1} \equiv \Phi^{0} = \Pi^{0}$
  &
  & \: $\beta^{0} =$ unknown                                                                             \\
\: $\gamma'^{0} \equiv \Sigma^{0} = \frac{N}{2}\eta^{ i j}  F_{ij} +  \partial_{i} \Pi^{i} $ \:
  &
  & \: $\lambda_i = e^2\eta_{li} \varphi^{l} $                                                                            \\

  & \: $\chi^0 \equiv \phi^{0} = \pi^{0}$
  & \: $\lambda_0  =unknow$                                                \\

  & \: $\chi^{i} \equiv \Phi^{i} = \pi^{i}$
  & \: $\beta_i  = -2Ne^{2}\eta_{ij} \varphi^{i} - \partial_i \varphi^0$                                                  \\

  & \: $\chi'^{i} \equiv \phi^{i} = \Pi^{i} -  (\eta^{ij} \varphi_{j} + N \eta^{ij}A_{j} )$
  & \:                                                   \\

  & \: $\chi'^{0} \equiv  \Lambda^{0} = e^2 \varphi^{0} + \frac{1}{2} \eta^{ij}F_{ij}$
  & \:                          \\

  & \: 
  & \:                         \\

  & \:  \:
  & \:                                                                                    \\
\hline
\end{tabular}\\ \\
Therefore,  we procedure to calculate the physical degrees of freedom as follows; there are 12 canonical variables, two first class constraints and six  second class constraints, thus,  there is one physical  degree of freedom as expected.\\
On the other hand, in order to obtain the extended action we must determine the unknown Lagrange multipliers. For this aim we will use the matrix whose entries are the Poisson brackets among the second class constraints, namely   $C_{\alpha \beta}$, thus  the  Lagrange multipliers will be determined by using  \cite{17, 17a}
\begin{equation}
\label{eq15}
\bar{u}^{\alpha} (y) \approx - \int \mathrm{d}^{2} x \, [ H (y) , \chi_{\alpha} (x) ] C^{-1}_{\alpha \beta}
(x,y) \, ,
\end{equation}
where $C^{-1}_{\alpha \beta}$ is such that
\begin{equation*}
\int \mathrm{d}^{2} z \, C_{\alpha \gamma} (x,z) \, C^{-1}_{\gamma \beta} (z,y) = \delta_{\alpha}^{\beta} \,
\delta^2(x-y) \,.
\end{equation*}
Hence, the  $C_{\alpha \beta}$ matrix is given by
\begin{equation*}
\label{eq:C_alpha_beta}
C_{\alpha \beta} = [\chi_{\alpha},\chi_{\beta}] =
\left(
  \begin{array}{cccccc}
     0                & 0            &   0    &  0  & 0 &   e^{2}                               \\
    0               &   0              &   0    &  0  &  -1  & 0                               \\
     0                &   0              &   0    &  1  &  0  &  0                                          \\
     0                &0              & -1    &  0  &  -2N  &   \partial_{(x) 2}                                              \\
     0                & 1              &   0    &  2N &  0  &   -\partial_{(x) 1}                                              \\
    -e^2 & 0     & 0 & -\partial_{(x) 2}  & \partial_{(x) 1}  &   0                                              \\
  \end{array}
\right) \delta^2(x-y) \,, 
%\label{eq15x}
\end{equation*}
and  calculating  the inverse  matrix $C^{-1}_{\alpha \beta}\equiv C^{\alpha \beta}$ turns out to be
\begin{equation}
\label{eq16}
C^{\alpha \beta} =
\left(
  \begin{array}{cccccc}
      0 & -\frac{\partial_{(x) 1}}{e^2}  &   -\frac{\partial_{(x) 2}}{e^2}                      &    0                      &   0                     &    -\frac{1}{e^2}              \\
     \frac{\partial_{(x) 1}}{e^2} &  0  & -2N                     &   0                       & 1                   &    0             \\
       \frac{\partial_{(x) 2}}{e^2}&  2N  &   0                     & -1    & 0            & 0       \\
      0 &  0  &1         &    0                     & 0                     &    0             \\
    0&  -1  & 0 &  0                     &   0                      &    0             \\
        \frac{1}{e^2} &  0  & 0                  &    0                     &   0                      &    0             \\
  \end{array}
\right) \delta^2(x-y) \,.
\end{equation}
In this manner, by using (\ref{eq15}) and (\ref{eq16})  the Lagrange  multipliers are given by 
\begin{IEEEeqnarray}{rCl}
 \beta_0 & = & \partial_{ i} \varphi^{i},         \yn \\
   \label{eql1}                                                      
   \beta_{i} & = & -2Ne^2 \eta_{ij}\varphi^j- \partial_i \varphi^{0},            \yn \\
   \label{eql2}                                                      
 \lambda_0 & = & 0,     \yn \\
   \label{eql3}                                                      
 \lambda_i&=&   e^2\eta_{ji}\varphi^{j},                \yn   
  \label{eql4}                                                      
\end{IEEEeqnarray}
therefore, all Lagrange's multipliers have been determined. \\
With all those results at hand, we are able to   identify the Dirac bracket  among two functionals on the phase space, namely $F , G$,   by means of 
\begin{equation}
\label{eq:Dirac}
\big\{ F(x) , G(y) \big\}_D = \big[ F(x) , G(y) \big]_P - \int \mathrm{d} z \mathrm{d} z' \big[ F(x) , \chi_{\alpha}
                            (z) \big] C^{\alpha \beta}(z,z')\big[ \chi_{\beta} (z') , G(y) \big], 
\end{equation}
where $\big[ F(x) , G(y) \big]_P$ is the Poisson bracket among the functionals $F$ and $G$,  $\chi_{\alpha}$ are the second class constraints and $C^{\alpha \beta}$ is given by
(\ref{eq16}). In this way,  the Dirac's brackets of the dynamical
variables $(Q_A \longrightarrow A_{\mu}$, $\varphi_{\mu})$ and  $(P^A \longrightarrow \Pi^{\mu}, \pi^{\mu})$ are given by  \\
$\big\{ Q_A (x) , Q_B (y) \big\}_D =  $ 
\begin{equation*}
\label{eq:dirac_Q_Q}
\bordermatrix{
                & A_{0} (y) & A_{1} (y) & A_{2} (y) & \varphi_{0} (y) & \varphi_{1} (y)        & \varphi_{2} (y) \cr
A_{0} (x)       & 0         & 0         & 0         & 0               & 0                      & 0               \cr
A_{1} (x)       & 0         & 0         & 0         & 0               & 0                      & 1               \cr
A_{2} (x)       & 0         & 0         & 0         & 0               & -1                     & 0               \cr
\varphi_{0} (x) & 0         & 0         & 0         & 0 & \frac{ \partial_{(x) 1}}{e^2} & \frac{ \partial_{(x) 2}}{e^2}    \cr
\varphi_{1} (x) & 0         & 0         & 1         &  -\frac{\partial_{(x) 1}}{e^2} & 0             & 2 N           \cr
\varphi_{2} (x) & 0         & -1        & 0         & -\frac{ \partial_{(x) 2}}{e^2} & -2N            & 0               \cr}
 \delta^2(x-y)
\end{equation*}

$\big\{ Q_{A} (x) , P^{B} (y) \big\}_D = $
\begin{equation*}
\label{eq:dirac_Q_P}
\bordermatrix{
                &  \Pi^{0} (y) &  \Pi^{1} (y)              & \Pi^{2} (y)             &  \pi^{0} (y) & \pi^{1} (y)
                                                                                     &  \pi^{2} (y)              \cr
A_{0} (x)       &  1           &  0                        & 0                       &  0           &  0
                                                                                     &  0                        \cr
A_{1} (x)       &  0           &  1                        & 0                       &  0           &  0
                                                                                     &  0                        \cr
A_{2} (x)       &  0           &  0                        & 1                       &  0           &  0
                                                                                     &  0                        \cr
\varphi_{0} (x) &  0           &  -\frac{\partial_{(x) 2}}{e^2} &  \frac{\partial_{(x) 1}}{e^2} &  0           &  0
                                                                                     &  0                        \cr
\varphi_{1} (x) &  0           &  -N                       & 0                       &  0           &  0
                                                                                     &  0                        \cr
\varphi_{2} (x) &  0           &  0                        & - N                     &  0           &  0
                                                                                     & 0                         \cr}
\delta^2(x-y), 
\end{equation*}
and $\big\{ P^{A} (x) , P^{B} (y) \big\}_D = 0$.\\
Furthermore, the   identification of the constraints and the Lagrange multipliers  will allow us  to  identify the extended action. This step is important because we will able to identify the extend Hamiltonian,  that will be used in the quantization scheme  in forthcoming works.  In fact, by using the first class constraints,  the second class constraints, and the Lagrange multipliers  (17a-17d) we find that the extended action takes the following form

\begin{eqnarray}
\label{eq:S_E}
&S_{E}& \big[ \varphi_{\mu}, \pi^{\mu}, A_{\mu}, \Pi^{\mu}, u_\alpha,v_\beta \big]
    =       \int \mathrm{d}^{4} x \Big\{ \dot{\varphi}_{\mu}\pi^{\mu}+ \dot{A}_{\mu}\Pi^{\mu} - \mathcal{H} - u_{0} \Pi^0 \nonumber \\ 
   &-&  u'_0\left(  \partial_{i} \Pi^{i}+   \frac{N}{2}\eta^{ i j} F_{ij} \right) - v_0 \pi^0 - v_i\pi^i- v'_{0} \left( e^{2} \varphi_{0} +\frac{1}{2} \eta^{i j} F_{ij}\right)  \nonumber \\ 
   &-& v'_i\left( \Pi^{i} - \eta^{ i j} \left( \varphi_{j} + N A_{j}  \right) \right)  
              \Big\},       
              \label{eq19}                                                              
   \end{eqnarray}
   where $ \mathcal{H}$ is given by 
   \begin{eqnarray}
    \mathcal{H}&=&-\frac{1}{2} \varphi_0 \eta^{ij}F_{ij}- \frac{e^2}{2}\varphi_0 \phi^0 - A_0\left( \partial_i \Pi^i + \frac{N}{2} \eta^{ij}F_{ij} \right)  + \frac{e^2}{2} \varphi_i \varphi^i  \nonumber \\&-& \pi^0 \partial_i \varphi^i- 2N e^2 \pi^{i} \eta_{ij}\varphi^{j}- \pi^{i} \partial_i \varphi^0-e^2 \eta_{li} \Pi^i  \varphi^l + e^2N\varphi^l A_l,  
   \end{eqnarray}
    and $(u_0, u'_0)$,   $(v_{0}, v'_0, v_i, v'_i)$ are Lagrange multipliers enforcing the first and second class constraints respectively. Hence, from the extended action  we are able to identify the extended Hamiltonian given by 
    \begin{equation}
    \mathcal{H}_{E}= \mathcal{H} + u_{0} \Pi^0 +   u'_0\left(  \partial_{i} \Pi^{i}+   \frac{N}{2}\eta^{ i j} F_{ij} \right).
    \end{equation}
    It is  easy to observe  that  in the limit $e\rightarrow 0$,  the action   (\ref{eq19}) is reduced to a  topological theory, this means, the theory is  reduced to a theory with diffeomorphisms covariance and laking of degrees of freedom as  expected. \\
 On the other hand,  the equations of motion  obtained from the extended action read 
\begin{equation*}
\label{eq:eqs_mov}
\begin{array}{rlcl}
( \delta A_{0} )       : & \dot{\Pi}^{0}     & = &   \partial_i \Pi^i + \frac{N}{2} \eta^{ij}F_{ij} ,                    \\[0.1em]
( \delta A_{i} )       : & \dot{\Pi}^{i}     & = &
                \eta^{ i j} \partial_{j} \{ \varphi_{0} + N A_{0} \} +  \eta^{ij}\partial_jv'_0   + N\eta^{ij}\partial_j u'_0  + e^2N\varphi^i ,         \\[0.1em]
( \delta \varphi_{0} ) : & \dot{\pi}^{0}     & = &
               \partial_{i} \pi^{i}  - e^{2}v'_0 - e^2\varphi^0 + \frac{1}{2} \eta^{ij} F_{ij} ,
                                                   \\[0.1em]
( \delta \varphi_{i} ) : & \dot{\pi}^{i}     & = &
                e^{2} \varphi^{i} -\partial^i \pi^0- 2N e^{2} \eta^{ i j}\pi_{j} - \eta^{ij}v'_j  + e^2\eta^{il}\Pi_l -e^2NA^i,       \\[0.1em]
( \delta \Pi^{0} ) :     & \dot{A}_{0}       & = & u_0,                                            \\[0.1em]
( \delta \Pi^{i} ) :     & \dot{A}_{i}       & = &
                 \partial_{i} A_{0}  - \partial_{i}u'_0 + v'_i  -e^2 \eta_{li}\varphi^l,                                                                \\[0.1em]
( \delta \pi^{0} ) :     & \dot{\varphi}_{0} & = &
                -\partial_{i} \varphi^{i} + v_0,                                             \\[0.1em]
( \delta \pi^{i} ) :     & \dot{\varphi}_{i} & = &
              -v_i-2Ne^2\eta_{ij}\varphi^j- \partial_i \varphi^0.
\end{array}
\end{equation*}
Finally, we will use   Castellani's  algorithm in order to find the  gauge transformations by means of the generator  $G = \int d^{2} x ( \dot{\epsilon}
\gamma_{1} + \epsilon \gamma_2 )$. Hence,  the gauge transformations of the theory are given by
\begin{eqnarray*}
\label{eq:trans_norma}
\delta_{0} \varphi_{\mu} (x) & = &  0,                                               \\[0.2em]
\delta_{0} \pi^{\mu} (x)     & = &  0,                                                   \\[0.2em]
\delta_{0} A_{\mu} (x)       & = &   \partial_{\mu} \epsilon,                               \\[0.2em]
\delta_{0} \Pi^{\mu} (x)     & = & \eta^{ i \mu} N \partial_{i} \epsilon,
\end{eqnarray*}
thus, the fields transform as
\begin{eqnarray}
\label{eq:trans_norma}
\nonumber 
\varphi'_{\mu} & \longrightarrow & \varphi_{\mu},     \nonumber                                                          \\[0.2em]
\pi'^{\mu}     & \longrightarrow & \pi^{\mu},                                     \nonumber                                 \\[0.2em]
A'_{\mu}       & \longrightarrow & A_{\mu} + \partial_{\mu} \epsilon,               \nonumber                              \\[0.2em]
\Pi'^{\mu}     & \longrightarrow & \Pi^{\mu} + N \eta^{ i \mu}  \partial_{i} \epsilon.
\end{eqnarray}
It is important to remark  that  these results have not been  reported in the literature. It is straightforward to show that the action (\ref{eq1}) is invariant under the transformations (\ref{eq:trans_norma}). On the other hand,  we are able to observe that  (\ref{eq:trans_norma}) are $U(1)$ gauge transformation of the connection field. In fact, the first class constraint $\gamma'^{0} $ is the generator of $U(1)$ transformations. \\
  In this manner, we have stablished all  necessary tools to perform the quantization of the theory. Our approach has been performed by following all the steps of  Dirac's formulation, we think that this   approach must be applied to other field theories just like general relativity in order to obtain control on its   constraints. Furthermore, a detailed hamiltonian analysis is the best guideline   to perform progress   of  quantization. 
\section{Hamiltonian dynamics for  non-abelian theory}
With the results obtained above,  it is possible to generalize them  for a non-abelian theory. We will  perform the Hamiltonian analysis by using the  $SU(N)$  group, hence,  the action  (\ref{eq1}) which turns on to be 
\begin{equation}
\label{eq22}
S[A_\mu^a,\varphi_\nu ^a] = \int_M \mathrm{d}^3 x \left\{ \frac{1}{2} \varepsilon^{\mu \nu \rho}
    \varphi_{a \mu} F^a_{\nu \rho} + \frac{1}{2} e^{2} \varphi_{a \mu} \varphi^{a \mu} + \frac{1}{2} N
    \varepsilon^{\mu \nu \rho} A^a_{\mu} \left[  F_{ a\nu \rho}- \frac{1}{3} f{_{abc}}A^b_{\nu}A^c_{\rho} \right] \right\} \,.
\end{equation}
where the manifold $M$, the constants $e$ and  $N$ have been defined above.  Here,  $F_{\mu \nu}^a=  \partial_{\mu} A^a_{\nu} -\partial_{\nu} A^a_{\mu} + f^{a}{_{bc}}A^b_{\mu}A^c_\nu$ is the curvature of $SU(N)$ and $a, b, c$ corresponds to be $SU(N)$ index. We have commented above, the analysis of the  action (\ref{eq1}) is an  important step   because it  could provide possible extensions to the analysis of  quantum dynamics of gravity in three dimensions,  hence,   our study  becomes to be mandatory in order to make a contribution for an   eventual quantum interpretation of the theory (\ref{eq22}).\\
 So, we will present the relevant results emerged form the analysis.   The canonical analysis of the action (\ref{eq22}) yields  the following extended action
\begin{eqnarray}
S_E[\varphi_{\mu}^{a}, \pi^{\mu}_{a}, A_\mu^{a}, \Pi_a^{\mu}, u^a_0,u^a, v_a^0,v^a_{i},v'^a_i,v^a] &=& \int dx^2 dt \Big\{  \dot{ \varphi}_{\mu}^{a} \pi^{\mu}_{a}  +\dot{A}_\mu^{a} \Pi_a^{\mu} - H- u^a_0 \gamma_a^{0}- u^a \gamma_a\nonumber \\ &-& 
 v^a_{0} \chi_a^{0}- v'^a_i \chi'^i_a -v^a_i \chi_a^i-v^a \chi_a  \Big \}, 
\end{eqnarray}
where $H $ is given by 
 \begin{eqnarray}
    \mathcal{H}&=&-\frac{1}{2} \varphi_{a0} \eta^{ij}F^a_{ij}- \frac{e^2}{2}\varphi_{a0} \phi^{a0} - A^a_0\left( D_i \Pi_a^i + \frac{N}{2} \eta^{ij}(\partial_i A_{aj}-\partial_j A_{ai}) +f_{ab}{^{c}}\varphi_i^b \pi^i_c - f_{ab}{^{c}} \varphi^{0b}\pi_{c}^{0} \right) \nonumber \\& +&  \frac{e^2}{2} \varphi_{ai} \varphi^{ai}  - \pi^0_a D_i \varphi^{ai}- 2N e^2 \pi_a^{i} \eta_{ij}\varphi^{aj}- \pi_a^{i} D_i \varphi^{0a}-e^2 \eta_{li} \Pi_a^i  \varphi^{al} +e^2 N\varphi^l_a A^a_l ,  
   \end{eqnarray}
$u^a_0, u^a$ are Lagrange multipliers enforcing  the following  first class constraints 
\begin{eqnarray}
\gamma^0_a &=&  \Pi^0_a \approx 0, \nonumber \\
\gamma_a  &=& D_i \Pi^i_a + \frac{N}{2}\eta^{ij} F_{aij} - f_{ab}{^{c}} \varphi^{0b}\pi_{c}^{0}+ f_{ab}{^{c}} \varphi^{b}_{i} \pi^i_c \approx 0. 
\end{eqnarray}
We would to comment,  that the constraint   $\gamma_a$ is identified as the Gauss constraint for the theory,  and  it is clear that its complete structure has not been reported in the literature. This step is  important because the correct identification  of the constraints is the best guideline to perform the quantization. On the other hand,   
$v_a^0, v^a_{i}, v'^a_i, v^a$ are Lagrange multipliers enforcing the following second class constraints 
\begin{eqnarray}
\chi^0_a &=& \pi^0_a \approx 0, \nonumber \\
\chi^i_a&=& \pi^i_a\approx 0, \nonumber \\ 
\chi'^i_a&=& \Pi^i_a -\eta^{ij} \varphi_{aj}- N\eta^{ij} A_{aj} \approx 0, \nonumber \\ 
\chi_a&=&  \frac{1}{2} \eta^{ij} F_{aij} + \varphi^0_a e^2  \approx 0. 
\end{eqnarray}
Now we calculate the constraints algebra,  the  non zero Poisson brackets among the constraints are given by 
\begin{eqnarray}
\{ \chi^i_a(x), \chi'^j_b(y) \} &=& - \eta^{ij} \sigma_{ab} \delta^2(x-y), \nonumber \\ 
\{ \chi_a(x), \chi'^j_b(y) \} &=& - \sigma_{ab} \eta^{li} \partial_l \delta^2(x-y) - \eta^{ik} f_{abg}A^g_k\delta^2(x-y), \nonumber \\ 
\{ \chi_a(x), \gamma_b(y) \} &=& -f_{ab}{^{g}}  \chi_g(x) \delta^2 (x-y) \approx 0,  \nonumber \\ 
\{ \gamma_a(x), \gamma_b(y) \} &=& f_{ab}{^{c}}  \gamma_c  (x) \delta^2(x-y)\approx 0 , \nonumber \\
\{ \gamma_a(x), \chi'^i_b(y) \} &=&  f_{ab}{^{c}}  \chi'^i_c(x) \delta^2(x-y) \approx 0, \nonumber \\
\{ \gamma_a(x), \chi^i_b(y) \} &=& - f_{ab}{^{c}} \chi^i_c(x) \delta^2(x-y) \approx 0, \nonumber\\
\{ \chi'^i_a(x), \chi'^j_b(y) \} &=& -2N \eta^{ij} \sigma_{ab} \delta^{2}(x-y), 
\label{eq29}
\end{eqnarray}
where $\sigma_{ab}$ is the metric of $SU(N)$. We  observe from Poisson's brackets that the algebra is closed. In fact,   $\gamma^0_a, \gamma_a $ are first class, while  $\chi^0_a,\chi^i_a, \chi'^i_a, \chi_a$ are second class constraints as is expected. \\
All those results allowing us  to perform the  counting of physical degrees of freedom as follows: There are $12(N^2-1)$ canonical variables, $2(N^2-1)$  first class constraints and $6(N^2-1)$ second class constraints, thus the physical degrees of freedom are  $(N^2-1)$.\\
From  the constraint algebra (\ref{eq29}), we are able to   identify the Dirac brackets for the theory,  by   observing   that  the  matrix whose elements are only the Poisson brackets among   second class constraints  is given by 
\begin{equation}
C_{\alpha\beta} = \left(
\begin{array}{rrrr}
0 \qquad &0 \qquad&0\qquad& \varphi^0_a  \\
  0 \qquad & 0 \qquad &- \eta^{ij} \sigma_{ab} \delta^2(x-y) \qquad & 0  \\
    0 \qquad & \eta^{ij} \sigma_{ab} \delta^2(x-y) \qquad & -2N \eta^{ij} \sigma_{ab} \delta^{2}(x-y) \qquad &  \sigma_{ab} \eta^{li} \partial_l \delta^2(x-y) - \eta^{ik} f_{abg}A^g_k\delta^2(x-y)  \\
-\varphi^0_a \qquad &0 \qquad &0\qquad & 0  \\ \label{eqa}
\end{array}
\right).
\end{equation}
In this manner,  the Dirac bracket among    two functionals $A$, $B$  is expressed   by 
\begin{equation}
\{A(x),B(y) \}_D= \{A(x),B(y)\}_P + \int du dv \{A(x), \zeta^\alpha(u) \} C^{-1}_{\alpha \beta}(u,v) \{\zeta^\beta(v), B(y) \}, 
\label{eq27}
\end{equation}
where $ \{A(x),B(y)\}_P$ is the usual Poisson bracket between the functionals $A,B$,   $\zeta^\alpha(u)=(\chi{_{IJ}}^{a}, \chi{_{IJ}}^{ab} ) $,  with   $C^{-1}_{\alpha \beta}(u,v)$  being   the inverse of (\ref{eqa}) which has a  trivial form. It is well known that Dirac's  bracket (\ref{eq27}) will be an essential ingredient   to make progress in the quantization of the theory \cite{12, 17, 17a}.\\
\newline
\section{ Concluding remarks}
In  this paper, we have performed a complete  Hamiltonian analysis of  abelian and non-abelian topological extensions of YM in three dimensions purposed by Govaerts. Our work  was developed by following all steps of Dirac's formulation without ignoring  some of them;  our approach allowed us to  report the full structure of the constraints,  the extended action and the extended Hamiltonian, all these results were not reported in \cite{7}. Hence, we have established the bases to perform the quantization of the theory by means canonical or path-integral formulations; we need to remember that the correct identification of the constraints,  the extended action and Dirac's brackets,   is the best guideline to perform the study of quantization and the identification  of observables of the theory. In this respect,  our approach allowed us identify all the Lagrange multipliers and then we could construct  the Dirac's brackets, thus, we are able to study the observables of the theory,  also this step  was not reported in \cite{7}. Our results are generic and can be extended for  a four-dimensional theory, in fact, in four dimensions the topological generalizations of YM could provide  generalized QCD theories and they could be amenable to test, therefore, our results also   contribute  to  the particle phenomenology.   We finish this paper with some remarks; with the present work, we have at hand a better classical description of the theories studied,  thus the approach developed along the paper is an alternative way to perform  a pure Hamiltonian framework for any theory under study. We are able to observe   alternative approaches in \cite{11b, 14},  where in the former    Dirac's study for Palatini  and the later for  Pleba\'nski  theories  were performed;  however those studies were not a pure Hamiltonian analysis as the present work. In this sense,  we expect that our approach will be an good alternative way to study the symmetries of  Plebanski action,  expecting to obtain a better description.  Therefore, the approach developed along this paper could be useful in that direction, all those ideas are in progress an will be reported in forthcoming works \cite{13}.
\newline
\newline
\noindent \textbf{Acknowledgements}\\[1ex]
This work was supported by CONACyT under grant CB-2010/157641. I  would like to thank to Gabriel Salom\'on L\'opez  and Luis Enrique Fern\'andez  for the
hospitality and friendship that they have offered me.

\end{document}